\documentclass[preprint,12pt]{elsarticle}

\usepackage{amssymb}
\usepackage{natbib,graphicx,setspace,lscape,longtable}
\usepackage{epsfig,graphicx}
\usepackage{extarrows}
\usepackage{lscape}
\usepackage{subfigure}
\usepackage{geometry}
\usepackage{amsmath,amsthm,amssymb,color}
\usepackage{rotating}
\usepackage{booktabs}
\usepackage{tabularx}
\usepackage{epstopdf}
\usepackage{geometry}
\usepackage{rotating}
\usepackage[colorlinks, citecolor=blue,linkcolor=blue]{hyperref}
\usepackage{bm}
\usepackage{threeparttable}
\usepackage{multirow}
\usepackage[center,compact]{titlesec}
\usepackage{tikz}
\usetikzlibrary{arrows.meta, positioning, fit}

\theoremstyle{remark}
\newtheorem{remark}{Remark}
\numberwithin{equation}{section}

\journal{Research Policy}

\begin{document}

\begin{frontmatter}



\title{\bf\Large LLM-powered Real-time Patent Citation Recommendation for Financial Technologies}


\author[aff1]{Tianang Deng\fnref{fn1}} 
\author[aff2]{Yu Deng\fnref{fn1}}
\author[aff3]{Tianchen Gao\corref{cor1}\fnref{fn1}} 
\author[aff1]{Yonghong Hu\fnref{fn1}}
\author[aff1]{Rui Pan\corref{cor1}\fnref{fn1}}

\affiliation[aff1]{organization={School of Statistics and Mathematics, Central University of Finance and Economics},
            addressline={39 South College Road}, 
            city={Beijing},
            postcode={100081}, 
            state={Beijing},
            country={China}}

\affiliation[aff2]{organization={Harbin Huiwen JetCreate Artificial Intelligence Technology Co., Ltd.},
            addressline={288 Zhigu Avenue}, 
            city={Harbin},
            postcode={150020}, 
            state={Heilongjiang},
            country={China}}

\affiliation[aff3]{organization={Beijing International Center for Mathematical Research (BICMR), Peking University},
            addressline={5 Yiheyuan Road}, 
            city={Beijing},
            postcode={100871}, 
            state={Beijing},
            country={China}}

\cortext[cor1]{Corresponding author. Email: gaotc@pku.edu.cn; ruipan@cufe.edu.cn}
\fntext[fn1]{The authors are listed in alphabetical order and contributed equally to this work.}

\begin{abstract}
Rapid financial innovation has been accompanied by a sharp increase in patenting activity, making timely and comprehensive prior-art discovery more difficult. This problem is especially evident in financial technologies, where innovations develop quickly, patent collections grow continuously, and citation recommendation systems must be updated as new applications arrive. Existing patent retrieval and citation recommendation methods typically rely on static indexes or periodic retraining, which limits their ability to operate effectively in such dynamic settings.
In this study, we propose a real-time patent citation recommendation framework designed for large and fast-changing financial patent corpora. Using a dataset of 428,843 financial patents granted by the China National Intellectual Property Administration (CNIPA) between 2000 and 2024 (IPC G06Q 10/00–50/00), we build a three-stage recommendation pipeline. The pipeline uses large language model (LLM) embeddings to represent the semantic content of patent abstracts, applies efficient approximate nearest-neighbor search to construct a manageable candidate set, and ranks candidates by semantic similarity to produce top-$k$ citation recommendations.
In addition to improving recommendation accuracy, the proposed framework directly addresses the dynamic nature of patent systems. By using an incremental indexing strategy based on hierarchical navigable small-world (HNSW) graphs, newly issued patents can be added without rebuilding the entire index. A rolling day-by-day update experiment shows that incremental updating improves recall while substantially reducing computational cost compared with rebuild-based indexing. The proposed method also consistently outperforms traditional text-based baselines and alternative nearest-neighbor retrieval approaches.
Overall, our results show that combining LLM-based semantic representations with scalable and update-efficient retrieval methods can effectively support prior-art discovery in rapidly evolving technological fields. From a policy and management perspective, the proposed framework provides a practical tool for patent examiners and firms to improve the completeness and timeliness of patent citations, thereby supporting higher-quality patent examination and better-informed innovation decisions in financial technologies.
\end{abstract}


\end{frontmatter}



\section{Introduction}
\label{intro}
Financial innovation is increasingly driven by digital technologies such as data analytics, machine learning, distributed ledgers, and platform-based business models, which reshape the provision of payments, credit, investment, and risk management services \citep{lee2018fintech, kou2025fintech}. Against this background, patents have become a key vehicle for codifying financial technology (FinTech) innovations and for measuring the technological capabilities of banks and other financial institutions \citep{chen2019valuable,zhao2022riding}. 
In contrast to industries with high technical barriers, such as pharmaceuticals or semiconductors, financial technologies frequently  build on existing algorithms \citep{kirkpatrick2017overcoming}, business models \citep{heston1993closed}, or software methods \citep{jakobsson1999proofs}. As a result, patent iteration occurs at a much faster pace, and even a delay of several months in identifying emerging ideas can lead to substantial losses, such as missed R\&D opportunities, avoidable litigation risks, or slower retrieval of relevant prior art during examination \citep{gans2008impact, charles2021examination}. 
These considerations underscore the importance of developing real-time patent recommendation systems that can keep pace with the rapid evolution of financial technologies.

Patent recommendation covers several related tasks, including patent citation recommendation for prior-art search \citep{lu2024knowledge}, similar-patent recommendation for retrieval \citep{xiao2023patent}, and patent recommendation for technology transfer or transaction decisions \citep{chen2023interpretable}. Among these tasks, patent citation recommendation has attracted particular attention because patent citations are widely regarded as the most direct measure of technological relatedness between patents \citep{chen2017patent}.
In practice, this attention is further reinforced by the fact that patent citations are often identified and curated during the examination process as prior-art references, commonly referred to as comparison documents. Accordingly, patent citation recommendation systems can accurately measure technological similarity among financial patents and recommend high-quality citations for a given query patent \citep{lu2024knowledge}. Such systems can help firms reduce redundant innovation and monitor competitors more effectively \citep{zhang2019patent, cappelli2023technological}, while also assisting examiners in conducting reliable novelty and non-obviousness assessments by surfacing the most relevant comparison documents \citep{kim2017examination}. However, existing patent citation recommendation systems face dual challenges of large-scale data and real-time updating, as evidenced by the rapid growth and large volume of China National Intellectual Property Administration (CNIPA) financial patent applications shown in Figure~\ref{fig:patnum}.
Therefore, practical systems should support real-time querying and incremental updating over large-scale patent collections, enabling both corporate users and examiners to capture emerging financial innovations promptly without sacrificing retrieval accuracy.

Patent citation recommendation is closely related to patent similarity measurement because most pipelines treat citation retrieval as a similarity-driven ranking problem, where the system ranks prior patents that are technologically related to a query patent and thus likely to serve as citable prior art \citep{chen2017patent}. In the literature, patent similarity is commonly operationalized using text-based similarity from patent content representations \citep{hain2022text}, classification-based similarity from International Patent Classification (IPC) or Cooperative Patent Classification (CPC) code overlap or hierarchical distance \citep{chen2011ipc}, and hybrid similarity that combines textual and structured signals \citep{choi2022two}.
Notably, most text-based similarity work in patent retrieval and recommendation has relied on conventional natural language processing (NLP) techniques, whereas the use of large language models (LLM) in patent-related tasks remains under-explored \citep{jiang2025natural}.
In addition, existing recommendation largely rely on periodic batch processing, which cannot scale with the explosive growth of patent data. Traditional patent similarity systems are typically rebuilt offline, requiring full re-embedding of documents or complete reconstruction of nearest-neighbor indices each time new patents arrive \citep{hain2022text,ascione2024comparative}. Hybrid text–classification models and citation-based methods face similar limitations: they depend on static corpora and are unable to incorporate newly filed applications without costly recomputation \citep{lu2024knowledge, fu2015patent, ali2024innovating}. 
This environment creates an urgent demand for systems capable of incrementally updating patent knowledge structures, supporting low-latency retrieval, and continuously surfacing the newest and most influential innovation signals, which motivates the real-time, large-scale patent citation recommendation system developed in this paper.

Building on these foundations, we propose a three-stage patent citation recommendation approach developed for the financial domain and designed to support real-time updates. In the first stage, we use an LLM-based embedding model, namely text-embedding-3-large model proposed by OpenAI, to encode patent abstracts into high-dimensional semantic vectors, allowing newly filed patents to be embedded and incorporated immediately. In the second stage, we apply an efficient approximate nearest neighbor search algorithm, Hierarchical Navigable Small World graph (HNSW) \citep{malkov2018efficient}, which naturally supports incremental graph expansion and real-time updates, enabling rapid integration of newly filed patents into the search structure without costly global recomputation. In the final stage, for a given query patent, we compute the cosine similarity between its semantic vector and those of the real-time updated candidate pool, rank candidates in descending order, and select the top-$k$ results as citation recommendations according to the specific application needs.

Our main contributions are threefold. First, we systematically collect, clean, and curate CNIPA patent data in the financial domain. We obtain 428,843 patents with the main IPC group ranging from G06Q 10/00 to G06Q 50/00, with application years ranging from 2000 to 2024. This corpus provides a comprehensive and representative benchmark dataset for studying Chinese financial patents. Second, we propose a scalable three-stage framework for large-scale patent citation recommendation in the financial domain. The framework uses an LLM-based embedding model to represent patent abstracts, applies HNSW to retrieve a tractable candidate pool, and then ranks candidates to generate the final recommendation list. Third, our framework supports real-time updates in continuously growing patent collections. By leveraging the incremental updating property of HNSW, newly filed patents can be embedded and inserted into the index without costly global rebuilding, which enables the system to keep pace with rapid patent turnover in finance and to continuously capture the latest technological developments that static, batch-updated methods may overlook.

The remainder of this paper is organized as follows. Section \ref{literature} reviews the related literature. Section \ref{data} introduces the data and provides descriptive analysis. Section \ref{approach} presents our patent citation recommendation approach. Section \ref{results} illustrates the result of our approach. Section \ref{incremental} demonstrates the performance of our method under incremental updates. Section \ref{conclusion} concludes the paper. 

\section{Literature Review}
\label{literature}
This section reviews related work along two dimensions, namely patent citation recommendation approaches, and domain characteristics of financial patents. We highlight how existing studies address each dimension and identify the gap in jointly satisfying domain specificity, large-scale efficiency, and real-time updates.

\subsection{Patent Citation Recommendation Approaches}
\label{patentcitation}

While general patent citation recommendation systems address retrieval or classification tasks across broad domains, patent citation recommendation focuses more specifically on suggesting technically relevant prior patents for a given application. This task has attracted increasing attention due to the growing size of patent databases and the need for precise prior-art discovery in support of novelty assessment and infringement risk analysis. Three main lines of research have emerged in the literature: methods based on textual similarity, methods using patent classification codes, and hybrid approaches. Among these, text similarity-based models have become especially prominent. These methods represent patents based on their textual content, including titles, abstracts, claims, or full texts, and compare them using metrics such as cosine similarity or semantic distance. Early approaches employed TF-IDF vectorization or simple keyword matching to identify related patents \citep{verma2011exploring}. As NLP tools advanced, distributed representations such as Word2Vec and Doc2Vec were applied, enabling the capture of deeper semantic connections across documents \citep{helmers2019automating}. Building on these developments, embedding-based methods leverage such textual representations to efficiently compute pairwise similarity across large-scale collections \citep{chen2020exploiting}. More recently, transformer-based models like BERT have been adapted to patent corpora, with domain-specific variants such as PatentSBERTa \citep{bekamiri2024patentsberta} and PaECTER \citep{ghosh2024paecter} outperforming traditional representations. Although transformer encoders are now widely used for patent text modeling, the broader use of large language models as a central component in patent citation recommendation has not yet become mainstream \citep{jiang2025natural}. These models encode richer contextual information, which is especially useful in fields like financial technology, where standard classification codes such as G06Q cover broad and overlapping areas. In such domains, textual methods can reveal subtle inventive distinctions that classification-based filters may overlook \citep{arts2018text}, and text-derived NLP measures provide additional fine-grained signals beyond classification-based indicators \citep{arts2021natural}.

Classification-based approaches instead rely on structured metadata like IPC or CPC codes to infer technological proximity. By measuring code overlap or semantic distance in the classification hierarchy, these models can filter candidate citations efficiently \citep{chen2011ipc}. However, because patent classes are often too coarse, classification-based methods may miss relevant documents that use different codes but share underlying concepts. As a result, their standalone effectiveness remains limited, particularly in emerging fields with fast-evolving taxonomies. Hybrid models attempt to combine the strengths of textual and structural signals. For instance, \cite{choi2022two} introduced a deep learning system that jointly encodes patent text and CPC classifications, leading to higher-quality recommendations. Others have framed the citation task as link prediction in patent citation networks or constructed knowledge graphs where text, metadata, and inventor-assignee relations form a heterogeneous node structure \citep{chen2023interpretable}. In parallel, graph-based reasoning systems such as PatentMind decompose similarity into multiple aspects, including technical features, application domains, and claim scope, yielding results that align more closely with expert assessments \citep{lu2024knowledge, yoo2025patentmindmultiaspectreasoninggraph}. These graph-based and hybrid approaches can capture citation patterns and domain co-evolution more comprehensively, but they often require large training corpora and complex infrastructure to deploy at scale.

Scalability is another critical challenge for patent citation recommendation because practical systems must search over rapidly growing corpora and return high-quality candidates efficiently. Performing dense similarity comparisons between a new patent and millions of existing ones is computationally prohibitive. To address this, many systems adopt $K$-Approximate Nearest Neighbors Search ($K$-ANNS) search methods, which reduce retrieval latency by using indexing techniques like locality-sensitive hashing (LSH), $k$-$d$ trees, or HNSW graphs. For example, \cite{hain2022text} built an $K$-ANNS-based infrastructure to construct a patent similarity network across the global corpus, allowing fast retrieval with minimal sacrifice in accuracy. Despite their speed, traditional $K$-ANNS algorithms struggle with dynamic updates. Most are designed for static corpora, meaning any addition of new patents often requires index rebuilding, which can be costly. Thus, building real-time patent citation recommendation systems remains a tradeoff between retrieval efficiency and update flexibility.

\subsection{Domain Characteristics for Financial Patents}
\label{domain}

Financial-patent citation recommendation lies at the intersection of patent analytics, patent similarity measurement, and recommender systems. Previous studies generally build recommendations by exploiting structured metadata such as classification codes, textual content embeddings, citation networks, or knowledge graphs. While these ingredients are widely applicable, the financial domain exhibits distinctive patenting and citation patterns that can weaken generic similarity heuristics and make timely patent citation recommendation particularly challenging. Accordingly, this subsection reviews domain-specific evidence from financial patent studies and synthesizes methodological lessons for building patent citation recommendation systems tailored to financial patents.

A growing literature uses patents to study FinTech and financial business-method innovation, showing that financial inventions are heavily concentrated in a small set of IPC/CPC areas such as G06Q and are frequently co-classified across closely related codes \citep{lee2017identifying}. This concentration implies that financial patents form dense neighborhoods in the classification space where many inventions share identical or near-identical code assignments \citep{lerner2002does}. Although classification-based similarity is useful for coarse filtering, it often lacks sufficient granularity to separate closely competing inventions in finance. In particular, IPC-based overlap or hierarchical distance can yield weak discrimination when the relevant subclasses are highly overlapping and densely populated, which makes it difficult to rank truly relevant prior art for a given query patent \citep{hall2009financial}. Consistent with broader evidence that classification-only proximity may be less precise than text-based similarity within dense technological areas \citep{arts2018text}, this limitation is especially salient for financial patents, where innovations are frequently incremental recombinations of algorithms, software, and business processes that can share the same top-level classes while differing materially in novelty and scope \citep{lerner2002does}. Moreover, prior work suggests that financial and business-method patents raise particular concerns about boundary clarity and enforceability, partly because claims can be broad, overlapping, and tightly tied to software or organizational innovations \citep{lerner2002does}.

In such settings, missing closely related prior art can translate into higher downstream uncertainty, including elevated dispute or litigation risk and less consistent examination outcomes \citep{hall2009financial}. This strengthens the practical value of patent citation recommendation systems that can surface highly relevant comparison documents for both firms and patent examiners in a timely manner. Despite advances in representation learning and retrieval, most existing systems are still designed around static corpora or slow refresh cycles. Building on the approaches reviewed in Section~\ref{patentcitation}, text-based and hybrid similarity models are generally better suited than pure classification filters for financial patents because they can capture fine-grained semantics within dense clusters \citep{hain2022text, arts2018text}.

A key practical constraint for deploying patent citation recommendation in finance is timeliness. Many existing pipelines perform model training and index maintenance offline on periodic snapshots \citep{hain2022text, bekamiri2024patentsberta}. As a consequence, incorporating newly filed patents often requires re-encoding large portions of the corpus or rebuilding nearest-neighbor indices, which requires substantial computational resources and makes frequent updates difficult at scale. This batch-oriented design is less suitable for financial patents, where the corpus grows continuously and users require up-to-date prior-art candidates \citep{lee2017identifying, lerner2002does}. This gap motivates practical patent citation recommendation systems that maintain retrieval accuracy while enabling low-latency querying and real-time incremental updates over large-scale patent collections.

\section{Data and Descriptive Analysis}
\label{data}

Most patent documents consist of three main components: bibliographic data, claims, and full text \citep{benzineb2011automated}. Bibliographic data typically includes the patent publication number, title, inventors, application date, IPC Symbol, and abstract. As a crucial type of textual data in patent documents, the abstract summarizes the technical features of the patent concisely, minimizing the inclusion of irrelevant words that could introduce noise \citep{li2018deeppatent,tshitoyan2019unsupervised}. Apart from the abstract, claims and full text constitute the other two types of textual data in patent documents. Due to their different purposes, their content and writing style also differ notably. Usually written in legal terms, the claim also describes the patent's technical features. However, it mainly focuses on distinguishing the patent from others for legal protection rather than highlighting their similarity, making them unsuitable for patent citation recommendation \citep{hain2022text, li2018deeppatent}. Obviously longer than abstract and claims, the full text description of the patent covers the most comprehensive information about the patent, containing the research background, technical details, and potential applications. But it can also introduce more noise \citep{noh2015keyword}. Furthermore, the large volume of text greatly increases the computational cost in the following research. 

Therefore, we select the abstract as the textual data used in our study. We obtain 428,843 patents with the main IPC group ranging from G06Q 10/00 to G06Q 50/00, which belong to subclass G06Q and represent patents related to finance from the database of CNIPA, with application years ranging from 2000 to 2024. For each patent, we collect the patent publication number, title, inventors, year of filing, IPC Symbols including main IPC Symbols and abstract. Table \ref{tab:example} shows an example patent in our dataset. It's worth noting that the text in our original data is all in Chinese. However, for readability, we present the English version of the patent as provided on Google Patents.

\begin{table}[ht]
\caption{An example to show 6 features we collected of a patent, including patent publication number, IPC symbols, title, year of filing, inventors and abstract.}
\label{tab:example}
\begin{center}
\resizebox{\textwidth}{!}{
\begin{tabular}{ccc}
\toprule
\textbf{Patent publication number} & \textbf{IPC symbols} & \textbf{Title}\\ 
\midrule
CN116258571A & G06Q40/03 & Loan inspection processing system and processing method \\
\midrule
\textbf{Year of filing} & \textbf{Inventors} & \textbf{Abstract}\\ 
\midrule
2023 & Lin Jiongcheng & \begin{tabular}[c]{@{}c@{}}The invention belongs to the technical field of loan inspection, and discloses \\ a loan inspection processing method, which comprises the following steps: ...  \end{tabular}\\
\bottomrule
\end{tabular}}
\end{center}
\end{table}

\begin{figure}[!ht]
    \centering
    \includegraphics[width=0.68\textwidth]{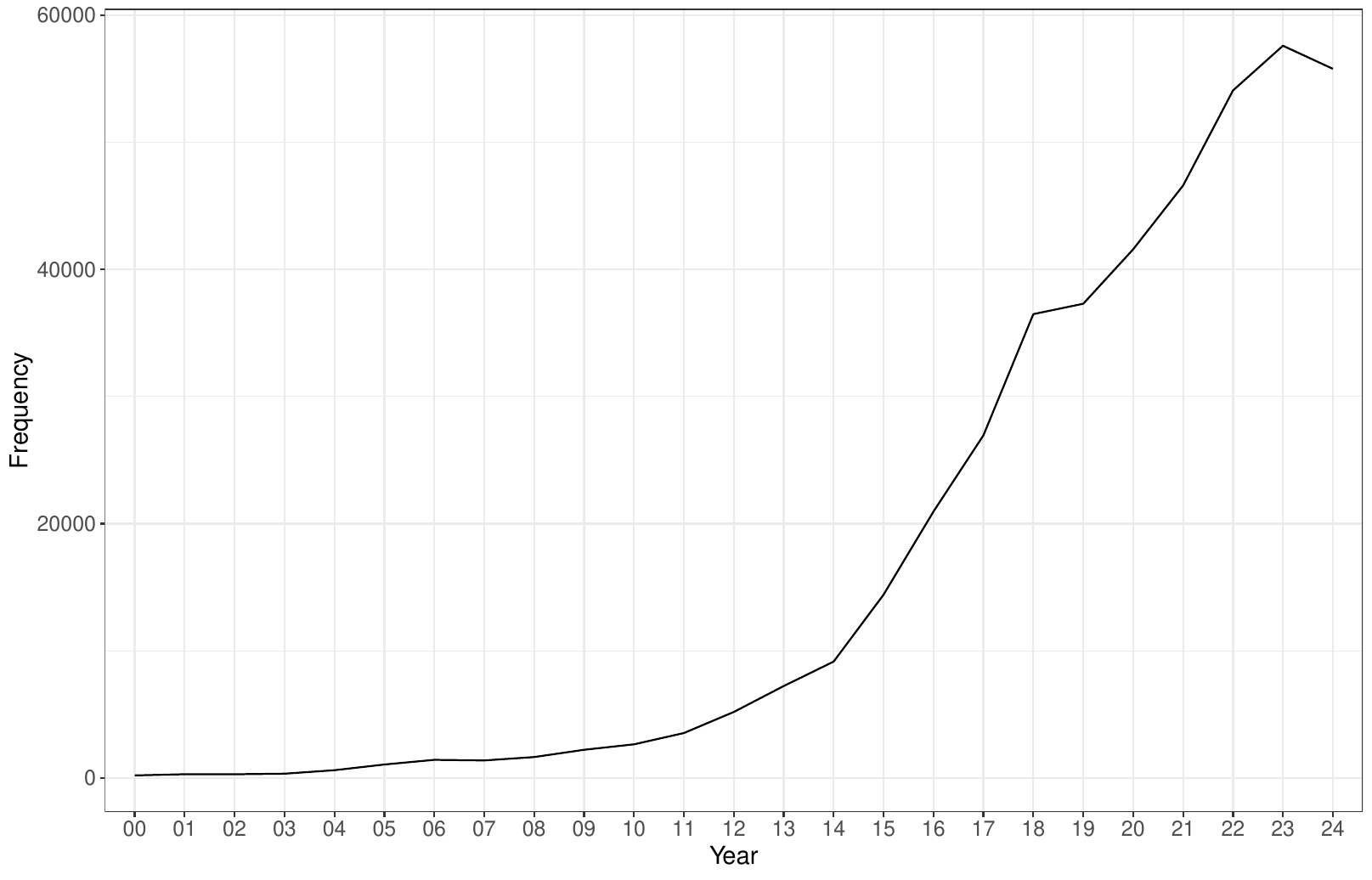}
    \caption{The annual number of published patents in the field of Finance applied to the CNIPA.}
    \label{fig:patnum}
\end{figure}

Figure \ref{fig:patnum} illustrates the annual number of published patents in the field of Finance applied to the CNIPA. The number of newly filed patents showed a gradual increase until 2008, followed by a rapid rise over the subsequent 15 years. Driven by technological innovation and policy support, the growth in patent filings is not limited to the field of Finance but extends across various sectors in China, resulting a substantial overall increase in patent numbers and making China the country with the largest number of patent applications in the world \citep{hu2017china, prud2019chinese}. This surge has increased the difficulty of patent examination, even reducing the quality of patents \citep{lu2024knowledge}. Therefore, it is essential to develop a method to recommend similar patents to examiners as potential citations. According to the regulations of the CNIPA, patents are published only after a maximum of 18 months from the application date. Consequently, some patents filed in the last two years have not yet been disclosed, resulting in a decline in the observed trend in 2024. In addition, we also use all patents filed in 2024 to evaluate the effectiveness of the patent citation recommendation method.

The abstract is a brief summary of the patent, typically providing a concise overview of its core technical content, which assists researchers in quickly understanding the technical information of the patent. In this section, we will present a descriptive analysis of the abstracts and conduct a preliminary exploration of the main research topics in financial patents. Since the abstracts under study are written in Chinese, it is important to note that, unlike many commonly used word-based languages such as English, a key feature of Chinese is its character-based nature. Since most Chinese characters do not carry complete meanings on their own, specific meanings are conveyed through the formation of words. Without delimiters such as spaces to separate words, Chinese sentences are composed in a continuous flow. Therefore, to analyze Chinese patent abstracts, the first step is to segment sentences into words. We utilize \textit{Jieba}, a
Chinese text segmentation tool, to segment the abstracts. 

Building on the segmented words, we extract keyphrases with a TextRank algorithm implemented in the Python package \textit{textrank4zh}. We restrict the candidate vocabulary to content words after segmentation and then construct an undirected, weighted word co-occurrence network within each abstract using a sliding window of size 10. Any two candidate words that appear within the same window are connected by an edge, and the edge weight accumulates their co-occurrence counts. In this network, we compute TextRank scores for all candidate words and these words are ordered by score and the top ten are retained as the keyword set $\mathbf{S}$. Keywords are formed directly from the original word sequence by merging adjacent words whenever both words belong to $\mathbf{S}$. We finally extract all the keywords from abstracts and compute their frequencies and visualize these frequencies with a word cloud, in which font size is proportional to the frequency of each keywords.

\begin{figure}[ht]
    \centering
    \includegraphics[width=0.98\textwidth]{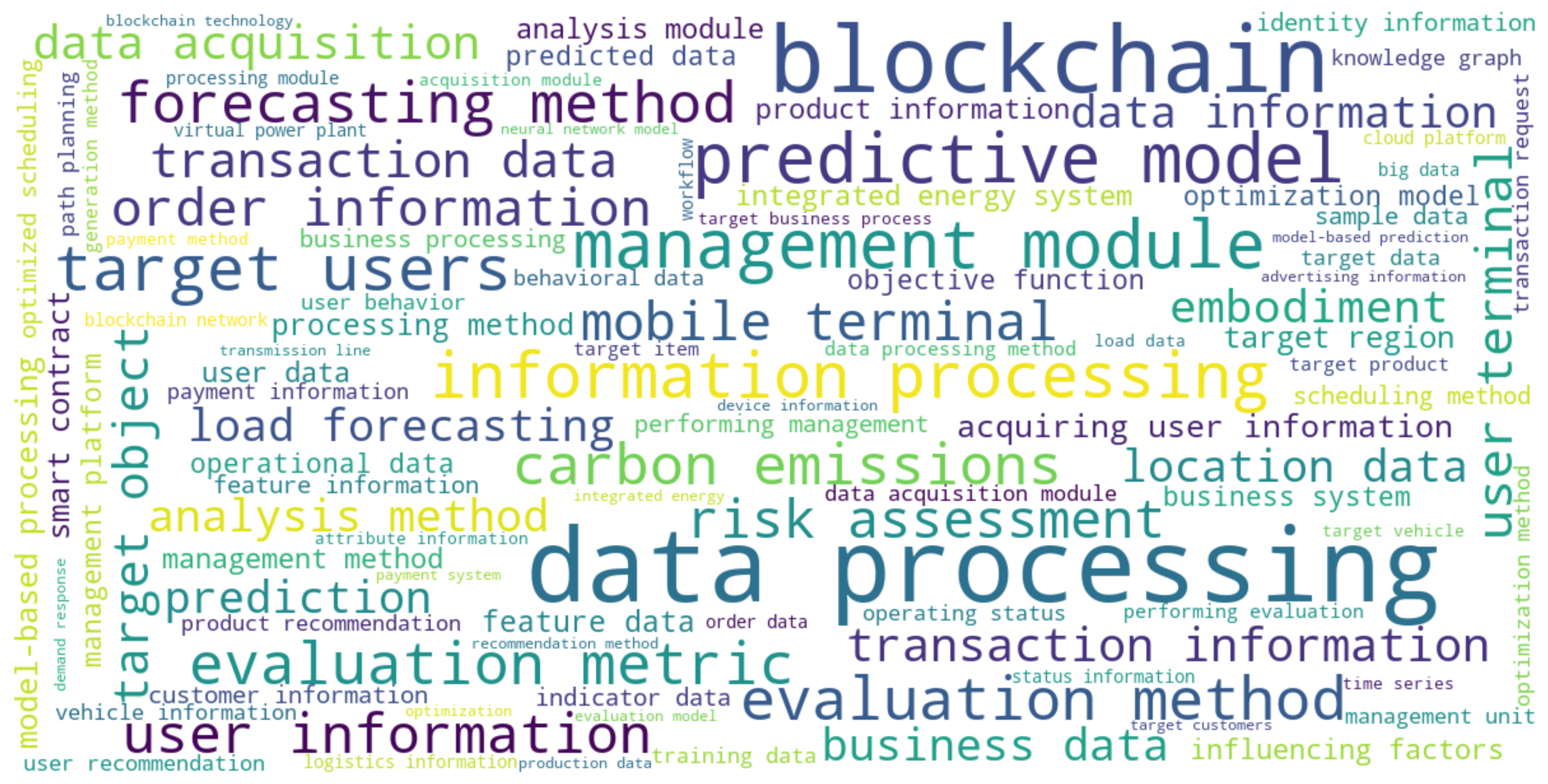}
    \caption{The word cloud of keywords in financial patent abstracts.}
    \label{fig:wordcloud}
\end{figure} 

Figure \ref{fig:wordcloud} presents the word cloud of keywords in financial patent abstracts. It illustrates a strong emphasis on data centric concepts, such as data processing, data acquisition, and information processing. Keywords including predictive model, optimization model, evaluation method, and neural network model suggest that many financial patents leverage machine learning approaches to analyze financial data. Core financial operations terminology such as transaction data, payment information, and order information also appears frequently. In addition, the high frequency of keywords such as carbon emissions and integrated energy system suggests sustained interest in the integration of finance and energy management.



\section{A Three-Stage Approach to Patent Citation Recommendation}
\label{approach}

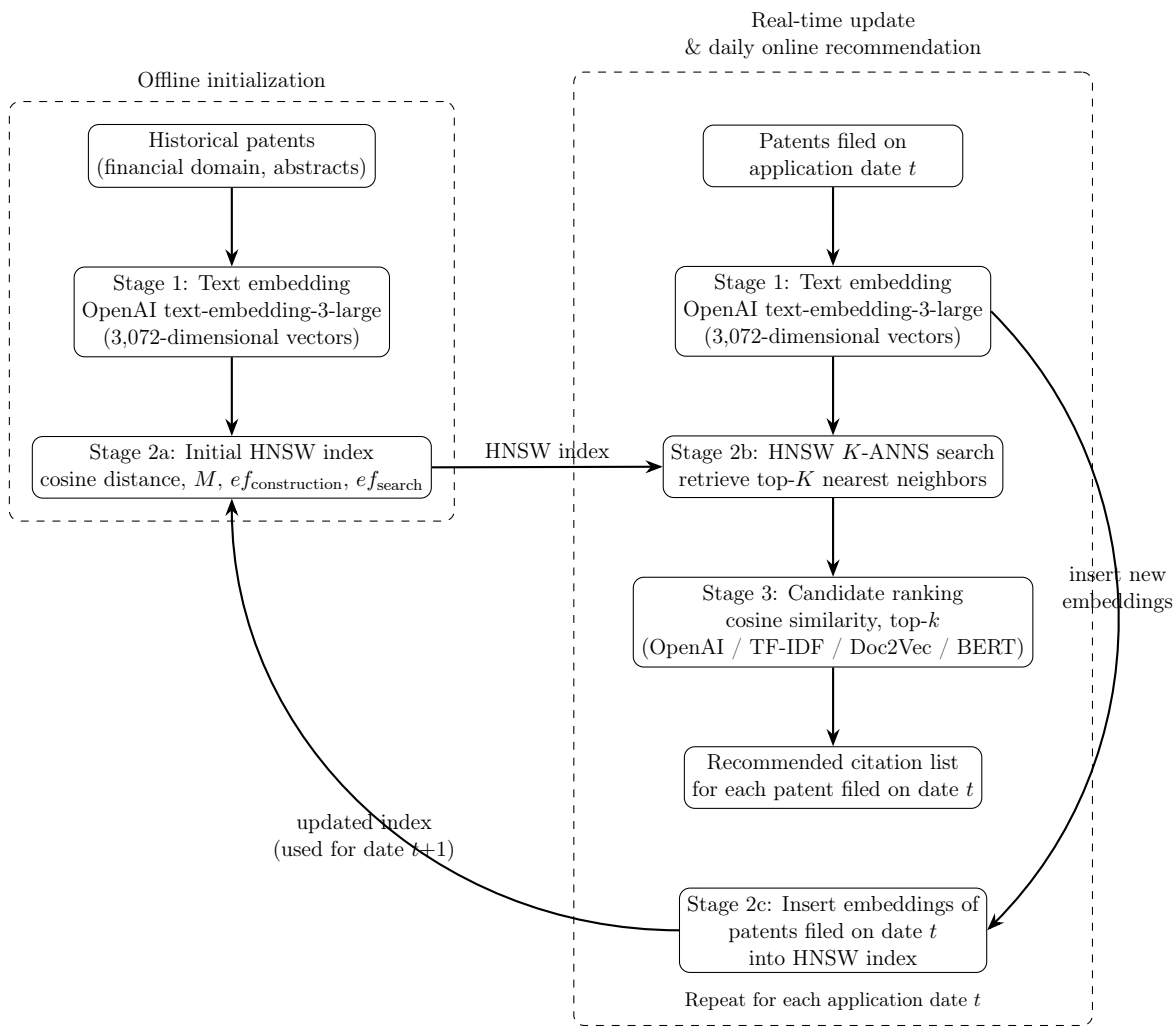
\begin{figure}[!ht]
    \centering
    \begin{tikzpicture}[
        scale=0.75,
        every node/.style={transform shape},
        node distance=1.4cm and 3.0cm,
        >=Stealth,
        font=\small,
        block/.style={
            rectangle,
            draw,
            rounded corners,
            align=center,
            minimum width=4.6cm,
            minimum height=1.0cm
        },
        line/.style={->, thick},
        group/.style={
            rectangle,
            rounded corners,
            draw,
            dashed,
            inner sep=0.3cm
        }
    ]

    \node[block] (hist) {Historical patents\\(financial domain, abstracts)};
    \node[block, below=of hist] (embed_hist) {Stage 1: Text embedding\\OpenAI text-embedding-3-large\\(3,072-dimensional vectors)};
    \node[block, below=of embed_hist] (hnsw_index) {Stage 2a: Initial HNSW index\\cosine distance, $M$, $ef_{\text{construction}}$, $ef_{\text{search}}$};

    \draw[line] (hist) -- (embed_hist);
    \draw[line] (embed_hist) -- (hnsw_index);

    \node[group, fit=(hist) (hnsw_index), label={[yshift=0.1cm]above:Offline initialization}] (offline_box) {};

    \node[block, right=5.8cm of hist] (day_input) {Patents filed on\\application date $t$};
    \node[block, below=of day_input] (embed_day) {Stage 1: Text embedding\\OpenAI text-embedding-3-large \\ (3,072-dimensional vectors)};
    \node[block, below=of embed_day] (hnsw_search) {Stage 2b: HNSW $K$-ANNS search\\retrieve top-$K$ nearest neighbors};
    \node[block, below=of hnsw_search] (ranking) {Stage 3: Candidate ranking\\cosine similarity, top-$k$\\(OpenAI / TF-IDF / Doc2Vec / BERT)};
    \node[block, below=of ranking] (output) {Recommended citation list\\for each patent filed on date $t$};
    \node[block, below=of output] (insert) {Stage 2c: Insert embeddings of\\patents filed on date $t$\\into HNSW index};

    \draw[line] (day_input) -- (embed_day);
    \draw[line] (embed_day) -- (hnsw_search);
    \draw[line] (hnsw_search) -- (ranking);
    \draw[line] (ranking) -- (output);
    \draw[line] (embed_day.east) to[out=-45, in=45]
    node[pos=0.5, above, align=center]{insert new \\embeddings}(insert.east);

    \node[
    group,
    fit=(day_input) (insert),
    label={[yshift=0.1cm, align=center]above:{Real-time update \\\& daily online recommendation}},
    inner xsep=1.4cm,  
    inner ysep=0.7cm   
] (online_box) {};

    \draw[line] (hnsw_index.east) -- node[above, align=center]{HNSW index} (hnsw_search.west);

    \draw[line] (insert.west) to[out=180, in=-90] node[below, align=center]{updated index\\(used for date $t{+}1$)} (hnsw_index.south);


    \node[below=0.2cm of insert] (loopnote) {\footnotesize Repeat for each application date $t$};

    \end{tikzpicture}    
    \caption{Flowchart of the proposed three-stage HNSW-based framework for patent citation recommendation, with the incremental real-time updating mechanism discussed in Section~\ref{incremental}.}

    \label{fig:flowchart}
\end{figure}

We propose a three-stage framework for large-scale patent citation recommendation. Figure \ref{fig:flowchart} illustrates the flowchart of our method. First, each patent abstract is embedded to a semantic vector using the text-embedding-3-large model proposed by OpenAI. Semantic proximity between patents is measured by cosine similarity. Second, to construct a tractable candidate set at scale and support real-time updates, we employ the HNSW, a graph-based $K$-ANNS, to retrieve the top-$K$ nearest neighbors for each query. In the final stage, we compute the cosine similarity between the query and each patent in the candidate pool, rank the candidates in descending order, and select the top-$k$  with the highest cosine similarity as the final recommendations.


\subsection{Text Embedding}
Text embedding is an important method in NLP that transforms discrete text data into continuous and fixed-dimensional numerical vectors. There are many approaches to producing text embeddings, including Transformer-based encoders, such as BERT \citep{devlin2019bert}, Sentence-BERT \citep{reimers2019sentence}, MPNet \citep{song2020mpnet}, dual-encoder retrieval models, including GTR-T5 \citep{ni2022large}, E5 \citep{wang2022text}, GTE \citep{li2023towards}, and classical neural methods such as Doc2Vec \citep{le2014distributed}. More recently, general-purpose embedding APIs have provided readily available representations for semantic retrieval, including text-embedding-3-small and text-embedding-3-large models proposed by OpenAI. Compared with traditional NLP pipelines, LLM-based embeddings can capture richer contextual semantics with minimal feature engineering, making them particularly effective for measuring fine-grained similarity in large and heterogeneous text corpora. In particular, text-embedding-3-large produces information-rich 3,072-dimensional vectors that capture fine-grained semantics and are well suited for large-scale similarity search in specialized corpora. The model is built on the Transformer architecture, whose self-attention mechanism enables effective contextual encoding of text sequences and improves semantic understanding \citep{vaswani2017attention}.

To measure the semantic similarity between patent abstracts, we utilize the text-embedding-3-large model to convert each abstract into a unit-length vector in $R^p$ with $p= 3,072$. Since semantically similar texts are mapped to nearby vectors, similarity between two patents can be measured directly in the embedding space. Specifically, following the recommendation of OpenAI, we compute cosine similarity, which is computationally efficient for unit-length vectors and yields the same ranking as Euclidean distance under normalization. For patent $A$ and patent $B$, let their embedding vectors be $v(A)$ and $v(B)$, then the cosine similarity is defined as $\text{sim}_{cos}(v(A), v(B)) = v(A)^{\text{T}}v(B)$, because $\|v(A)\|_2=\|v(B)\|_2=1$ for vectors output by text-embedding-3-large model.


\subsection{Candidate Pool Construction}
For relatively small datasets, we can compute the cosine similarity between the embedding vectors of each pair of patents. However, our dataset contains a large number of patents, and the volume of documents faced by examiners when searching for patent citations is considerably larger. In such cases, directly computing the pairwise cosine similarity between all patents leads to substantial computational cost. This highlights the need for a method that can rapidly identify neighboring observations in the embedding space for large-scale data. $K$-ANNS algorithms are designed to retrieve the neighbors close enough to the given data point within large-scale datasets, substantially reducing the computational time required for evaluating pairwise vector similarities \citep{tian2023approximate}. A variety of widely used $K$-ANNS search algorithms have been developed, such as ANNOY \citep{bernhardsson2017annoy}, LSH \citep{indyk1998approximate} and HNSW. In this study, we adopt the HNSW algorithm to construct the candidate pool due to its high recall, efficient search, and native support for incremental index updates that avoid costly global rebuilding when new patents arrive.

The idea of HNSW is to support approximate nearest neighbor search by organizing data points into a multi-layer navigable small-world graph. From an algorithmic perspective, HNSW consists of two main components, namely index construction and neighbor search \citep{malkov2018efficient}. During index construction, each layer is a proximity graph that connects each point to a small number of neighbors. 
Higher layers are more sparse and contain fewer points, while the bottom layer contains all data points. The construction process is incremental. When a new data point is inserted, it is assigned a maximum layer randomly according to an exponentially decaying distribution, which ensures that only a few points are placed on the top layers. Starting from the highest layer of the current graph, the new point performs a greedy search to identify an entry point at each layer until reaching its assigned maximum layer. From this layer down to the bottom layer, the point is included in the graph and establishes connections to a set of its nearest neighbors within each layer. This hierarchical insertion strategy enables the graph to simultaneously preserve long-range connections that facilitate efficient global navigation and short-range connections that enhance local refinement.

Given a query point, HNSW employs a hierarchical search strategy commonly referred to as the \textit{SEARCH-LAYER} algorithm. The search begins at the top layer, using a specific entry point, and performs a greedy search. The algorithm iteratively moves to the neighbor closest to the query until no closer neighbor can be found. The final position of this search is then used as the entry point for the next lower layer. This process is repeated layer by layer, progressively narrowing the search region. At the bottom layer, which contains all data points, the algorithm performs a more exhaustive local search within a candidate set, thereby identifying the $K$ nearest neighbors of the query point. These neighbors are subsequently returned as the approximate nearest neighbors.

In our study, we utilize the HNSW to construct the candidate pool by building an incrementally updated graph over all embedding vectors of patents filed between 2000 and 2023, which corresponds to Stage~2a in Figure~\ref{fig:flowchart}. We construct an HNSW index using cosine distance as the similarity metric, with each patent embedding represented as the 3,072-dimensional vector. The index is initialized with a maximum capacity equal to the number of embeddings, and key hyper-parameters were set to $M=48$, $ef_{construction}=100$, and $ef_{search}=2,000$, balancing recall accuracy and query efficiency during approximate nearest neighbor retrieval. Based on the constructed index, we then perform $K$-ANNS for each query patent to retrieve the top-$K$ nearest neighbors as candidates, which corresponds to Stage~2b in Figure~\ref{fig:flowchart}. In our implementation, for each patent filed in 2024, we set $K=1,000$ and use the retrieved neighbors as the candidate pool for citation recommendation.



\begin{remark}
HNSW builds a multi-layer proximity-graph index and supports a $K$-ANNS procedure that takes an arbitrary query point $q$, starts from the upper layer, and greedily descends to return the $K$ closest stored elements. Consequently, the nearest neighbors in the indexed corpus of a newly filed patent that was not used during index construction can be retrieved efficiently without retraining or rebuilding the index.
\end{remark}

\begin{remark}
The HNSW provides native support for incremental index construction and updating the elements. After an HNSW index is constructed, newly filed patents can be inserted directly by running the standard insertion routine without reconstructing the index. In contrast, some tree-based methods such as ANNOY are effectively static or require costly rebuilds to incorporate new data. This incremental property substantially reduces update latency and compute, enabling continuous ingestion and up-to-date retrieval in streaming settings,which aligns well with the workflow of patent examiners who require timely citation recommendations. The real-time incremental updating mechanism (Stage~2c) is discussed in Section~\ref{incremental}.
\end{remark}

\subsection{Candidate Ranking and Recommendation}
After constructing the candidate pools using HNSW, the final stage of our approach involves ranking the patents within each candidate pool to produce the final citation recommendations. For a given query patent, we compute cosine similarity between its embedding vector and those of the 1,000 candidates retrieved in the previous stage, and then select the top-$k$ as the final recommendations. The recommendation cutoff $k$ is configurable and can be increased or reduced to meet practical requirements.To enhance robustness and flexibility, this ranking procedure can be performed using different types of vector representations, such as embeddings generated by the text-embedding-3-large model, term frequency–inverse document frequency (TF-IDF) vectors, document vectors from the Doc2Vec model, and contextual embeddings produced by BERT.

\noindent \textbf{TF-IDF}: Term Frequency–Inverse Document Frequency (TF-IDF) is a classical vectorization method in information retrieval that assigns weights to terms according to their frequency within a document and their rarity across the corpus. This representation captures the importance of terms for distinguishing documents, and has been widely used in text similarity and retrieval tasks. In our study, we selected the 10,000 most frequent terms from the entire corpus and computed the TF-IDF values of each patent abstract with respect to these terms, thereby obtaining a 10,000-dimensional TF-IDF vector representation for each document.

\noindent \textbf{Doc2Vec}: The Doc2Vec model extends the Word2Vec framework by learning fixed-length vector representations for entire documents \citep{le2014distributed}. In our study, we trained a Doc2Vec model on the full corpus of patent abstracts using the distributed memory (PV-DM) architecture. The training was performed with a vector size of 500, a context window of 10, and a minimum word frequency threshold of 2. We employed negative sampling with 10 negative examples and applied subsampling with a threshold of $10^{-5}$ for frequent words. The model was trained for 50 epochs with 10 worker threads, and a fixed random seed was used to ensure reproducibility of results.

\noindent \textbf{BERT}: Bidirectional Encoder Representations from Transformers (BERT) is a pre-trained language model based on the Transformer architecture, which leverages bidirectional self-attention to capture contextual dependencies between words \citep{devlin2019bert}. In our study, we employed the \texttt{bert-base-chinese} model to encode patent abstracts. Each abstract was tokenized with a maximum sequence length of 500, and the encoded sequences were processed in batches of 300 samples. The model parameters were kept frozen, and the representation corresponding to the [CLS] token in the final hidden layer was extracted as the document embedding. This yielded a 768-dimensional contextual vector for each patent abstract.

\section{Results}
\label{results}

To evaluate the practical effectiveness of our patent citation recommendation framework on a large-scale financial patent corpus, we report empirical results under a time-ordered evaluation setting that mirrors real-world use. The retrieval index is built from earlier-filed patents and recommendations are generated for newly filed patents. We benchmark the proposed pipeline against alternative $K$-ANNS algorithm and widely used text-representation baselines, and we further compare our results with the similar-document function provided by Google Patents as an external reference. In addition, we present quantitative performance comparisons using standard ranking metrics including Mean Reciprocal Rank (MRR), Normalized Discounted Cumulative Gain (nDCG), and recall rates at different sizes of the recommendation set (Rec@$k$) in Section \ref{modelcompare}, and then provide a qualitative case study in Section \ref{example} to illustrate how different methods vary in the prior art they retrieve and the resulting recommendation lists.

\subsection{Model Comparison}
\label{modelcompare}
To evaluate the effectiveness of the proposed method, we compare its performance against another approximate nearest neighbors search algorithms, ANNOY, and several baseline models, including TF-IDF, Doc2Vec and BERT. Specifically, after embedding all patent abstracts through the text-embedding-3-large model, we construct an HNSW graph based on all financial patents filed before 2023. Subsequently, for a total of 15,733 financial patents filed in 2024 that have at least one comparison document in our dataset, we retrieve their $K=1,000$ nearest neighbors from the graph. 
These neighbors are then ranked in descending order by cosine similarity in the 3,072-dimensional embedding space, and the top-$k$ patents are returned as candidate citations. To make the evaluation more stable and to reflect practical needs at different recommendation scales, we report results at multiple cutoffs, setting $k\in\{10,50,100,200\}$.
Similarly, cosine similarity has also been calculated between each of the patents filed in 2024 mentioned above and their 1,000 nearest neighbors in the HNSW graph based on these vector representations obtained from baseline models including TF-IDF, Doc2Vec and other text embedding methods including BERT. The neighbors are then ranked accordingly. The top $k$ patents from each ranking will be selected as the candidate citations generated by these three baseline approaches and compared with our previous results.

As a comparison, we also construct an ANNOY forest containing 100 trees based on embedding vectors obtained from text-embedding-3-large model of all financial patents filed before 2023, and for each of the 15,733 financial patents filed in 2024 we retrieve its 1,000 approximate nearest neighbors from this forest, rank them in descending order according to the cosine similarity of their embedding vectors, and select the top $k$ patents as the candidate citations produced by the ANNOY-based approach. To evaluate their performance in citation recommendation, we select several common evaluation metrics, including MRR \citep{voorhees2000building}, nDCG \citep{jarvelin2002cumulated}, and Rec@$k$.


\begin{table}[ht]
\caption{Comparison results of different text vectorization models and nearest neighbor search methods.}
\label{tab:evaluation}
\footnotesize
\begin{center}
\resizebox{\textwidth}{!}{
\begin{tabular}{ccccccc}
\toprule
\textbf{Models} & \textbf{MRR} & \textbf{nDCG} & \textbf{Rec@10} & \textbf{Rec@50} & \textbf{Rec@100} & \textbf{Rec@200} \\
\midrule
Exact-Large & 0.1782 & 0.1831 & 0.1309 & 0.2512 & 0.3196 & 0.3914 \\
\textbf{HNSW-Large} & \textbf{0.1782} & \textbf{0.1830} & \textbf{0.1309} & \textbf{0.2511} & \textbf{0.3194} & \textbf{0.3912} \\
ANNOY-Lagre & 0.1775 & 0.1814 & 0.1301  & 0.2489 & 0.3160 & 0.3860 \\
HNSW-TF-IDF & 0.0786 & 0.0806 & 0.0617 & 0.1430 & 0.1912 & 0.2507 \\
HNSW-Doc2Vec & 0.0397 & 0.0493 & 0.0275 & 0.0753 & 0.1172 & 0.1806 \\
HNSW-BERT & 0.0348 & 0.0404 & 0.0234 & 0.0622 & 0.0936 & 0.1455 \\
\bottomrule
\end{tabular}
}
\end{center}
\end{table}

Table \ref{tab:evaluation} reports the comparative performance of various text vectorization models and nearest neighbor search methods. As shown, the Exact-Large model, which performs exhaustive nearest neighbor search based on cosine similarity, achieves the highest overall performance across all evaluation metrics. Its results serve as the upper-bound benchmark for retrieval accuracy. Notably, the proposed method that combines the text-embedding-3-large model with HNSW achieves the best performance among all the ANN-based models across all evaluation metrics. Specifically, HNSW-Large yields the highest values in MRR of 0.1782, nDCG of 0.1831, and recall across all cutoff points, indicating its superior ability to retrieve relevant citations in higher ranking positions. The performance of ANNOY-Large is only marginally lower suggesting that both HNSW and ANNOY are effective for large-scale patent retrieval, though HNSW offers a slight advantage in recall and ranking precision due to its hierarchical graph structure and improved local search efficiency. In contrast, traditional text representation methods such as TF-IDF and Doc2Vec exhibit substantially lower retrieval performance, with HNSW-TF-IDF achieving an MRR of only 0.0787 and HNSW-Doc2Vec decreasing further to 0.0406. These results demonstrate the limitations of shallow text representations in capturing complex semantic relations between patents. Similarly, HNSW-BERT shows inferior results compared with the proposed HNSW-Large, suggesting that while BERT embeddings encode contextual information, they may not align optimally with cosine similarity in high-dimensional semantic retrieval tasks without further fine-tuning.

As a comprehensive global patent search platform, Google Patents primarily serves as a large-scale database and search interface for worldwide patent information. In addition to basic search and browsing, it provides, for each patent, a list of up to 25 similar documents retrieved from its corpus based solely on text similarity \citep{google_patents_result_viewer}. Building on the comparative analysis of retrieval models in Table \ref{tab:evaluation}, we therefore treat the similar-document lists generated by Google Patents as an external benchmark for evaluating our proposed citation recommendation framework. 

Specifically, we first collect the similar documents on Google Patents for the 15,733 patents filed in 2024 mentioned above. Almost all patents have 25 similar documents, with only 45 patents having fewer than 25. Because the similar document lists may contain both scholarly articles and patents, whereas our study focuses exclusively on recommending patent citations, we retain only those similar documents that are patents, yielding on average 19.79 similar patents per focal patent. Moreover, as cited documents are typically filed earlier than the citing patent, and Google Patents’ similar document selection is based solely on text similarity without considering temporal order, we further filter the similar patents by requiring their filing date to precede that of the origin patent. After this temporal filtering, each patent has on average 13.70 earlier-filed similar patents. 

Because the retained similar patents are obtained from a global patent corpus spanning all technological fields, we first evaluate their performance by computing, for each focal patent filed in 2024, the coverage of its entire set of cited patents. Given that each patent has on average 13.70 similar patents after temporal filtering, we select the top 13 patents returned by our method for each focal patent as recommended citations, so that the recommendation list sizes are aligned. In addition, to assess performance in the financial domain, we further restrict the similar patents for each focal patent to those that appear in our dataset, which yields an average of 3.61 similar patents per patent. We then compute the coverage of these similar patents with respect to the cited patents of the focal patents within our dataset and set the recommendation list size of our method to 3 to ensure a balanced comparison.

\begin{table}[ht]
\caption{Recall comparison between Google Patents similar documents and our method with matched recommendation sizes, where Rec@13 matches the average list size of earlier-filed similar patents, 13.70 per query, and Rec@3 matches the average list size after restricting similar patents to our financial-domain dataset, 3.61 per query.}

\label{tab:google}
\begin{center}
\footnotesize
\begin{tabular}{ccc}
\toprule
\textbf{Models} & \textbf{Rec@3} & \textbf{Rec@13} \\
\midrule
HNSW-Large & 0.0712 & 0.1463 \\
Google Patents & 0.0175 & 0.0122 \\
\bottomrule
\end{tabular}
\end{center}
\end{table}

Table \ref{tab:google} illustrates the comparison results of Google Patents similar documents and our method. Whether we consider similar patents drawn from the full patent corpus or those restricted to the financial domain, their coverage of actual citations is below 2\%, and is lower than the coverage achieved by our method with recommendation lists of the same size. Furthermore, comparing the two sets of recommendations from Google Patents reveals that, when the recommendation scope is restricted to the financial domain, the coverage of financial-domain citations is slightly higher than the coverage of all citations obtained from the full patent corpus. This suggests that, when using text similarity to identify citation relationships between patents, focusing on the financial domain leads to better performance.Our method focuses on patent texts in the financial domain and leverages text embeddings generated by large language models, thereby achieving better citation recommendation performance for financial patents.

\subsection{An illustrative example}
\label{example}

After comparing the overall recommendation metrics of several methods on large-scale datasets, we further illustrate the differences in their recommendation content through a case study. Specifically, we select the patent “Parallel execution method, device, electronic device and program product for blockchain transactions” from the IPC main group G06Q 20 as a representative example. Table \ref{tab:top20comparison} presents the correctly retrieved patents and their ranks among the top-20 recommended results obtained using three vectorization methods, including text-embedding-3-large, TF-IDF, and BERT, under the HNSW approximate search setting. Since the Doc2Vec model failed to retrieve any actually cited patents within its top-20 recommendations, we instead report the highest-ranked true citation among its candidate results.

\begin{table}[!ht]
\renewcommand{\arraystretch}{1.8}
\caption{Comparison of correctly retrieved patents among the top 20 results recommended by each model.}
\label{tab:top20comparison}
\begin{center}
\resizebox{\textwidth}{!}{
\begin{tabular}{ccc}
\toprule
\textbf{Title of the query patent} & \textbf{Models} & \textbf{Rank and title of the correctly retrieved patents} \\
\midrule
\multirow{20}{*}{\begin{tabular}[c]{@{}c@{}}Parallel execution method, device, \\  electronic device and program \\ product for blockchain transactions \end{tabular}} & \multirow{8}{*}{text-embedding-3-large} & \begin{tabular}[c]{@{}c@{}} 2. Processing method and device for transactions in \\ block chain, electronic equipment and storage medium \end{tabular} \\
& & \begin{tabular}[c]{@{}c@{}} 6. Transaction execution method, device and storage \\ medium  based on DAG block chain system \end{tabular} \\
& & \begin{tabular}[c]{@{}c@{}} 9. Block chain parallel transaction method, device, \\ equipment and storage medium \end{tabular} \\
& & \begin{tabular}[c]{@{}c@{}} 16. Method for parallel processing, device, equipment \\ and the storage medium of block chain data \end{tabular} \\
\cline{2-3}
& \multirow{8}{*}{TF-IDF} & \begin{tabular}[c]{@{}c@{}} 1. Transaction execution method, device and storage \\ medium based on DAG block chain system \end{tabular} \\
& & \begin{tabular}[c]{@{}c@{}} 7. Block chain parallel transaction execution method \\ and device and electronic equipment \end{tabular} \\
& & \begin{tabular}[c]{@{}c@{}} 8. Method for parallel processing, device, equipment \\ and the storage medium of block chain data \end{tabular} \\
& & \begin{tabular}[c]{@{}c@{}} 20. Block chain transaction parallel execution method \\ and device based on associated semantics \end{tabular} \\
\cline{2-3}
& BERT & \begin{tabular}[c]{@{}c@{}} 6. Block chain parallel transaction method, device, \\ equipment and storage medium \end{tabular} \\
\cline{2-3}
& Doc2Vec & \begin{tabular}[c]{@{}c@{}} 101. Parallel execution of transactions in a distributed \\ ledger system \end{tabular} \\
\bottomrule
\end{tabular}
}
\end{center}
\end{table}

It can be observed that both the text-embedding-3-large and TF-IDF methods successfully identify 4 true citations within the top-20 recommendations, while the ranks of the correct recommendations produced by the former are consistently higher than those from the latter. This observation is consistent with the conclusions drawn in the previous subsection. Further analysis of the abstracts and full texts of these patents reveals that the first correct recommendation retrieved by text-embedding-3-large is highly similar to the query patent. They both determine dependency relationships among blockchain transactions and achieve parallel processing through hierarchical execution, without requiring the participation of a Directed Acyclic Graph (DAG) in the dependency discrimination process. In contrast, among the four high-ranked recommendations produced by the TF-IDF method, three are based on DAG-related approaches—particularly the patent “Transaction execution method, device and storage medium based on DAG blockchain system”, which ranks first among all candidates. This finding suggests that the TF-IDF method, relying solely on basic textual features, tends to capture textual similarity rather than technical relevance. By contrast, the text-embedding-3-large model can more comprehensively capture semantic information, thereby identifying patents that are technically more relevant and offering better assistance in patent examination tasks.

\section{Incremental Updates for Real-Time Citation Recommendation}
\label{incremental}
Unlike many ANN methods, including ANNOY, HNSW has a major advantage in that it supports continuous incremental index construction.
This means that once an HNSW graph has been constructed at a given point in time, newly filed patents can be directly inserted into the existing graph without reconstructing the index. In the practical work of patent examiners, they must assess the similarity between each newly filed patent and all the previously filed patents in order to ensure that no important references are overlooked. In practice, large numbers of new patent applications are filed almost every day, which implies that an ANN-based search framework must support continuous dynamic updates to efficiently and accurately provide examiners with candidate comparison documents. Consequently, this property of HNSW substantially reduces the time required for index reconstruction and makes the approach operationally feasible. 

To demonstrate both the improvement in recommendation performance brought by dynamic graph updates and the time savings achieved by HNSW relative to alternative methods, we conduct a day-by-day update experiment. Specifically, we first order the 15,733 patents filed in 2024 that have at least one comparison document in our dataset by their application date. After discarding several days on which no new patent applications occur, this yields 341 distinct application dates. We begin with an HNSW graph constructed from all patent embeddings for applications filed on or before 31 December 2023 and use this graph to recommend citations for all patents filed on the first application date in 2024. The embedding vectors of these patents are then inserted into the HNSW graph, which is subsequently used to generate citation recommendations for patents filed on the next application date, and this rolling-update procedure is repeated until all 341 application dates have been processed. As a comparison, we conduct a similar experiment using the ANNOY method, in which the ANNOY forest must be reconstructed for each application date in 2024. In addition, the static HNSW graph and ANNOY forest constructed from patents filed between 2000 and 2023 are also included as baselines. For this experiment, we compute the evaluation metrics reported in Section 5.1 and record the time record the time required to complete each procedure. 

\begin{table}[ht]
\caption{Performance and time-cost comparison of static HNSW/ANNOY, incremental HNSW and reconstructed ANNOY search methods for patent citation recommendation. TC denotes the time cost measured in seconds. The best result for each metric is highlighted in bold. Incremental HNSW achieves the best performance on almost all metrics.}
\label{tab:incremental}
\begin{center}
\resizebox{\textwidth}{!}{
\begin{tabular}{cccccccc}
\toprule
\textbf{Search Method} & \textbf{MRR} & \textbf{nDCG} & \textbf{Rec@10} & \textbf{Rec@50} & \textbf{Rec@100} & \textbf{Rec@200} & \textbf{TC (s)} \\
\midrule
Static HNSW & 0.1782 & 0.1830 & 0.1309 & 0.2511 & 0.3194 & 0.3912 & \textbf{859.6} \\
Static ANNOY & 0.1775 & 0.1814 & 0.1301  & 0.2489 & 0.3160 & 0.3860 & 1470.9\\
Reconstructed ANNOY & \textbf{0.1928} & 0.2040 & 0.1425 & 0.2777 & 0.3555 & 0.4385 & 49146.4 \\
\textbf{Incremental HNSW}  & 0.1926 & \textbf{0.2055} & \textbf{0.1433}  & \textbf{0.2801} & \textbf{0.3590} & \textbf{0.4443} & 1147.3 \\
\bottomrule
\end{tabular}
}
\end{center}
\end{table}

Table \ref{tab:incremental} illustrates the performance and time-cost comparison of static HNSW, incremental HNSW, static ANNOY, and reconstructed ANNOY search methods for patent citation recommendation. Incremental HNSW achieves the best performance on all recommendation metrics except MRR, where reconstructed ANNOY is slightly better. Compared with static HNSW, all recommendation metrics of incremental HNSW improve substantially. For example, the recall at 200 increases from 39.12\% to 44.44\%. In addition, the total time cost of incremental HNSW is only 288 seconds higher than that of static HNSW, which corresponds to an additional overhead of roughly 5 ms per newly filed patent, which is an acceptable cost in practical applications. In contrast, both static and reconstructed ANNOY yield lower recommendation performance than static and incremental HNSW, while incremental HNSW is even faster than static ANNOY and reconstructed ANNOY is more than 30 times slower than its static counterpart. These results demonstrate that the HNSW method not only achieves high recommendation accuracy, but also enables rapid graph updates in real-world deployment, making it feasible to dynamically update the recommendation scope on a daily basis.

\section{Conclusion}
\label{conclusion}
This study addresses a growing challenge in financial innovation systems: how to support timely and reliable prior-art discovery in the presence of rapidly expanding and continuously evolving patent corpora. Using a large-scale CNIPA financial patent dataset, we develop and evaluate a real-time patent citation recommendation framework that combines LLM-based semantic representations, efficient approximate nearest-neighbor retrieval, and incremental updating under a time-ordered evaluation setting. The results show that LLM-based embeddings substantially improve semantic matching over traditional text representations, while HNSW-based retrieval delivers higher recall than alternative indexing schemes with manageable computational cost. More importantly, the proposed framework enables real-time updating without full index reconstruction, allowing citation recommendations to remain aligned with newly issued patents in fast-moving technological domains.

Beyond its empirical performance, the key contribution of this study lies in demonstrating the feasibility and value of dynamic citation recommendation. Existing patent search and recommendation systems are often designed as static or periodically updated tools, implicitly assuming stable knowledge environments. Our findings suggest that such assumptions are increasingly misaligned with innovation dynamics in financial technologies, where both the volume and content of patented knowledge change rapidly. By explicitly incorporating incremental updating into the recommendation process, the proposed approach offers a practical pathway toward citation systems that better reflect the temporal structure of innovation.

From a policy and managerial perspective, these results have several implications. For patent offices, real-time citation recommendation can support examiners in identifying relevant prior art more comprehensively and at earlier stages of examination, potentially improving citation completeness and examination quality while reducing search burden. For firms and innovators, timely access to relevant prior patents may help clarify the competitive and technological landscape, reduce inadvertent infringement risks, and inform strategic R\&D and patenting decisions. More broadly, improved citation practices can enhance the informational value of patent data, benefiting empirical research and evidence-based innovation policy.

The study also opens several directions for future research. One natural extension is to incorporate richer patent content, such as claims and full descriptions, which may further improve legal relevance and scope matching. Another direction is to integrate additional signals—such as classification codes, citation-network structure, and temporal constraints—into the ranking stage, potentially through supervised learning-to-rank approaches. Finally, evaluating real-time citation recommendation systems in applied settings remains an important challenge. Future work could examine robustness across patent offices, explore alternative update frequencies, and investigate human-in-the-loop designs in which examiner feedback is used to refine recommendations and guide system updates over time.

\bibliographystyle{elsarticle-num} 
\bibliography{ref}

@article{lee2018fintech,
  title={Fintech: Ecosystem, business models, investment decisions, and challenges},
  author={Lee, In and Shin, Yong Jae},
  journal={Bus. Horiz.},
  volume={61},
  number={1},
  pages={35--46},
  year={2018},
  publisher={Elsevier},
  note={https://doi.org/10.1016/j.bushor.2017.09.003},
}

@article{kou2025fintech,
  title={{FinTech}: a literature review of emerging financial technologies and applications},
  author={Kou, Gang and Lu, Yang},
  journal={Financ. Innov.},
  volume={11},
  number={1},
  pages={1},
  year={2025},
  publisher={Springer},
  note={https://doi.org/10.1186/s40854-024-00668-6}
}

@article{chen2019valuable,
  title={How valuable is {FinTech} innovation?},
  author={Chen, Mark A and Wu, Qinxi and Yang, Baozhong},
  journal={Rev. Financ. Stud.},
  volume={32},
  number={5},
  pages={2062--2106},
  year={2019},
  publisher={Oxford University Press},
  note={https://doi.org/10.1093/rfs/hhy130}  
}

@article{zhao2022riding,
  title={Riding the {FinTech} innovation wave: {FinTech}, patents and bank performance},
  author={Zhao, Jinsong and Li, Xinghao and Yu, Chin-Hsien and Chen, Shi and Lee, Chi-Chuan},
  journal={J. Int. Money Finance},
  volume={122},
  pages={102552},
  year={2022},
  publisher={Elsevier},
  note={https://doi.org/10.1016/j.jimonfin.2021.102552}  
}

@article{kirkpatrick2017overcoming,
  title={Overcoming catastrophic forgetting in neural networks},
  author={Kirkpatrick, James and Pascanu, Razvan and Rabinowitz, Neil and Veness, Joel and Desjardins, Guillaume and Rusu, Andrei A and Milan, Kieran and Quan, John and Ramalho, Tiago and Grabska-Barwinska, Agnieszka and Hassabis, Demis and Clopath, Claudia and Kumaran, Dharshan and Hadsell, Raia},
  journal={Proc. Natl. Acad. Sci. U.S.A.},
  volume={114},
  number={13},
  pages={3521--3526},
  year={2017},
  publisher={National Academy of Sciences},
  note={https://doi.org/10.1073/pnas.1611835114}  
}

@article{heston1993closed,
  title={A closed-form solution for options with stochastic volatility with applications to bond and currency options},
  author={Heston, Steven L},
  journal={Rev. Financ. Stud.},
  volume={6},
  number={2},
  pages={327--343},
  year={1993},
  publisher={Oxford University Press},
  note={https://doi.org/10.1093/rfs/6.2.327}  
}

@incollection{jakobsson1999proofs,
  title={Proofs of Work and Bread Pudding Protocols (Extended Abstract)},
  author={Jakobsson, Markus and Juels, Ari},
  booktitle={Secure Information Networks},
  editor={Preneel, Bart},
  pages={258--272},
  publisher={Springer},
  address={Boston, MA},
  year={1999}
}

@article{gans2008impact,
  title={The impact of uncertain intellectual property rights on the market for ideas: Evidence from patent grant delays},
  author={Gans, Joshua S and Hsu, David H and Stern, Scott},
  journal={Manag. Sci.},
  volume={54},
  number={5},
  pages={982--997},
  year={2008},
  publisher={INFORMS},
  note={https://doi.org/10.1287/mnsc.1070.0814}  
}

@article{charles2021examination,
  title={Examination incentives, learning, and patent office outcomes: The use of examiner’s amendments at the {USPTO}},
  author={Charles, AW and Pairolero, Nicholas A and Teodorescu, Mike HM},
  journal={Res. Policy},
  volume={50},
  number={10},
  pages={104360},
  year={2021},
  publisher={Elsevier},
  note={https://doi.org/10.1016/j.respol.2021.104360}  
}

@article{lu2024knowledge,
  title={Knowledge graph enhanced citation recommendation model for patent examiners},
  author={Lu, Yonghe and Tong, Xinyu and Xiong, Xin and Zhu, Hou},
  journal={Scientometrics},
  volume={129},
  number={4},
  pages={2181--2203},
  year={2024},
  publisher={Springer},
  note={https://doi.org/10.1007/s11192-024-04966-9}  
}

@article{xiao2023patent,
  title={A patent recommendation method based on {KG} representation learning},
  author={Xiao, Yan and Li, Congdong and Th{\"u}rer, Matthias},
  journal={Eng. Appl. Artif. Intell.},
  volume={126},
  pages={106722},
  year={2023},
  publisher={Elsevier},
  note={https://doi.org/10.1016/j.engappai.2023.106722}  
}

@article{chen2023interpretable,
  title={Interpretable patent recommendation with knowledge graph and deep learning},
  author={Chen, Han and Deng, Weiwei},
  journal={Sci. Rep.},
  volume={13},
  number={1},
  pages={2586},
  year={2023},
  publisher={Nature Publishing Group UK London},
  note={https://doi.org/10.1038/s41598-023-28766-y}  
}

@article{chen2017patent,
  title={Do patent citations indicate knowledge linkage? {The} evidence from text similarities between patents and their citations},
  author={Chen, Lixin},
  journal={J. Informetr.},
  volume={11},
  number={1},
  pages={63--79},
  year={2017},
  publisher={Elsevier},
  note={https://doi.org/10.1016/j.joi.2016.04.018}  
}

@article{zhang2019patent,
  title={How patent signals affect venture capital: {The} evidence of bio-pharmaceutical start-ups in {China}},
  author={Zhang, Lili and Guo, Ying and Sun, Ganlu},
  journal={Technol. Forecast. Soc. Change},
  volume={145},
  pages={93--104},
  year={2019},
  publisher={Elsevier},
  note={https://doi.org/10.1016/j.techfore.2019.05.013}  
}

@article{cappelli2023technological,
  title={Technological competition and patent strategy: {Protecting} innovation, preempting rivals and defending the freedom to operate},
  author={Cappelli, Riccardo and Corsino, Marco and Laursen, Keld and Torrisi, Salvatore},
  journal={Res. Policy},
  volume={52},
  number={6},
  pages={104785},
  year={2023},
  publisher={Elsevier},
  note={https://doi.org/10.1016/j.respol.2023.104785}  
}

@article{kim2017examination,
  title={Examination workloads, grant decision bias and examination quality of patent office},
  author={Kim, Yee Kyoung and Oh, Jun Byoung},
  journal={Res. Policy},
  volume={46},
  number={5},
  pages={1005--1019},
  year={2017},
  publisher={Elsevier},
  note={https://doi.org/10.1016/j.respol.2017.03.007}  
}

@article{hain2022text,
  title={A text-embedding-based approach to measuring patent-to-patent technological similarity},
  author={Hain, Daniel S and Jurowetzki, Roman and Buchmann, Tobias and Wolf, Patrick},
  journal={Technol. Forecast. Soc. Change},
  volume={177},
  pages={121559},
  year={2022},
  publisher={Elsevier},
  note={https://doi.org/10.1016/j.techfore.2022.121559}  
}

@article{chen2011ipc,
  title={An {IPC}-based vector space model for patent retrieval},
  author={Chen, Yen-Liang and Chiu, Yu-Ting},
  journal={Inf. Process. Manage.},
  volume={47},
  number={3},
  pages={309--322},
  year={2011},
  publisher={Elsevier},
  note={https://doi.org/10.1016/j.ipm.2010.06.001}  
}

@article{choi2022two,
  title={A two-stage deep learning-based system for patent citation recommendation},
  author={Choi, Jaewoong and Lee, Jiho and Yoon, Janghyeok and Jang, Sion and Kim, Jaeyoung and Choi, Sungchul},
  journal={Scientometrics},
  volume={127},
  number={11},
  pages={6615--6636},
  year={2022},
  publisher={Springer},
  note={https://doi.org/10.1007/s11192-022-04301-0}  
}

@article{jiang2025natural,
  title={Natural language processing in the patent domain: a survey},
  author={Jiang, Lekang and Goetz, Stephan M},
  journal={Artif. Intell. Rev.},
  volume={58},
  number={7},
  pages={214},
  year={2025},
  publisher={Springer},
  note={https://doi.org/10.1007/s10462-025-11168-z}  
}

@article{ascione2024comparative,
  title={A comparative analysis of embedding models for patent similarity},
  author={Ascione, Grazia Sveva and Sterzi, Valerio},
  journal={arXiv preprint},
  year={2024},
  note={{,\space https://arxiv.org/abs/2403.16630}}
}

@inproceedings{fu2015patent,
  title={Patent citation recommendation for examiners},
  author={Fu, Tao-yang and Lei, Zhen and Lee, Wang-Chien},
  booktitle={Proc. IEEE ICDM},
  pages={751--756},
  year={2015},
  organization={IEEE},
  note={https://doi.org/10.1109/ICDM.2015.151}  
}

@article{ali2024innovating,
  title={Innovating patent retrieval: a comprehensive review of techniques, trends, and challenges in prior art searches},
  author={Ali, Amna and Tufail, Ali and De Silva, Liyanage Chandratilak and Abas, Pg Emeroylariffion},
  journal={Appl. Syst. Innov.},
  volume={7},
  number={5},
  pages={91},
  year={2024},
  publisher={MDPI},
  note={https://doi.org/10.3390/asi7050091}  
}

@article{malkov2018efficient,
  title={Efficient and robust approximate nearest neighbor search using hierarchical navigable small world graphs},
  author={Malkov, Yu A and Yashunin, Dmitry A},
  journal={IEEE Trans. Pattern Anal. Mach. Intell.},
  volume={42},
  number={4},
  pages={824--836},
  year={2018},
  publisher={IEEE},
  note={https://doi.org/10.1109/TPAMI.2018.2889473}  
}

@inproceedings{verma2011exploring,
  title={Exploring Keyphrase Extraction and {IPC} Classification Vectors for Prior Art Search},
  author={Verma, Manisha and Varma, Vasudeva},
  booktitle={Working Notes for CLEF 2011 Conference},
  editor={Petras, Vivien and Forner, Pamela and Clough, Paul and Ferro, Nicola},
  pages={19--22},  
  publisher={CEUR-WS.org},
address={Amsterdam, The Netherlands},
  year={2011}
}

@article{helmers2019automating,
  title={Automating the search for a patent’s prior art with a full text similarity search},
  author={Helmers, Lea and Horn, Franziska and Biegler, Franziska and Oppermann, Tim and M{\"u}ller, Klaus-Robert},
  journal={PloS one},
  volume={14},
  number={3},
  pages={e0212103},
  year={2019},
  publisher={Public Library of Science San Francisco, CA USA},
  note={https://doi.org/10.1371/journal.pone.0212103}  
}

@article{chen2020exploiting,
  title={Exploiting word embedding for heterogeneous topic model towards patent recommendation},
  author={Chen, Jie and Chen, Jialin and Zhao, Shu and Zhang, Yanping and Tang, Jie},
  journal={Scientometrics},
  volume={125},
  number={3},
  pages={2091--2108},
  year={2020},
  publisher={Springer},
  note={https://doi.org/10.1007/s11192-020-03666-4}  
}

@article{bekamiri2024patentsberta,
  title={{PatentSBERTa}: A deep {NLP} based hybrid model for patent distance and classification using augmented {SBERT}},
  author={Bekamiri, Hamid and Hain, Daniel S and Jurowetzki, Roman},
  journal={Technol. Forecast. Soc. Change},
  volume={206},
  pages={123536},
  year={2024},
  publisher={Elsevier},
  note={https://doi.org/10.1016/j.techfore.2024.123536}  
}

@misc{ghosh2024paecter,
  title={{PaECTER}: Patent-level representation learning using citation-informed transformers},
  author={Ghosh, Mainak and Rose, Michael E and Erhardt, Sebastian and Buunk, Erik and Harhoff, Dietmar},
  year={2024},
  journal={arXiv preprint},
  note = {{,\space https://arxiv.org/abs/2402.19411}}
}

@article{arts2018text,
  title={Text matching to measure patent similarity},
  author={Arts, Sam and Cassiman, Bruno and Gomez, Juan Carlos},
  journal={Strateg. Manage. J.},
  volume={39},
  number={1},
  pages={62--84},
  year={2018},
  publisher={Wiley Online Library},
  note={https://doi.org/10.1002/smj.2699}  
}

@article{arts2021natural,
  title={Natural language processing to identify the creation and impact of new technologies in patent text: {Code}, data, and new measures},
  author={Arts, Sam and Hou, Jianan and Gomez, Juan Carlos},
  journal={Res. Policy},
  volume={50},
  number={2},
  pages={104144},
  year={2021},
  publisher={Elsevier},
  note={https://doi.org/10.1016/j.respol.2020.104144}  
}

@article{yoo2025patentmindmultiaspectreasoninggraph,
  title={{PatentMind}: A Multi-Aspect Reasoning Graph for Patent Similarity Evaluation}, 
  author={Yongmin Yoo and Qiongkai Xu and Longbing Cao},
  year={2025},
  journal={arXiv preprint},
  note={{,\space https://arxiv.org/abs/2505.19347}}
}

@article{lee2017identifying,
  title={Identifying emerging trends of financial business method patents},
  author={Lee, Won Sang and Sohn, So Young},
  journal={Sustainability},
  volume={9},
  number={9},
  pages={1670},
  year={2017},
  publisher={MDPI},
  note={https://doi.org/10.3390/su9091670}  
}

@article{lerner2002does,
  title={Where does State Street lead? {A} first look at finance patents, 1971 to 2000},
  author={Lerner, Josh},
  journal={J. Finance},
  volume={57},
  number={2},
  pages={901--930},
  year={2002},
  publisher={Wiley Online Library},
  note={https://doi.org/10.1111/1540-6261.00446}  
}

@article{hall2009financial,
  title={Financial patenting in {Europe}},
  author={Hall, Bronwyn H and Thoma, Grid and Torrisi, Salvatore},
  journal={Eur. Manag. Rev.},
  volume={6},
  number={1},
  pages={45--63},
  year={2009},
  publisher={Wiley Online Library},
  note={https://doi.org/10.1057/emr.2009.3}  
}

@incollection{benzineb2011automated,
  title     = {Automated Patent Classification},
  author    = {Benzineb, Karim and Guyot, Jacques},
  booktitle = {Current Challenges in Patent Information Retrieval},
  editor    = {Lupu, Mihai and Mayer, Katja and Tait, John and Trippe, Anthony J.},
  pages     = {239--261},
  publisher = {Springer},
  address   = {Berlin, Heidelberg},
  year      = {2011}
}

@article{li2018deeppatent,
  title={{DeepPatent}: patent classification with convolutional neural networks and word embedding},
  author={Li, Shaobo and Hu, Jie and Cui, Yuxin and Hu, Jianjun},
  journal={Scientometrics},
  volume={117},
  number={2},
  pages={721--744},
  year={2018},
  publisher={Springer},
  note={https://doi.org/10.1007/s11192-018-2905-5}  
}

@article{tshitoyan2019unsupervised,
  title={Unsupervised word embeddings capture latent knowledge from materials science literature},
  author={Tshitoyan, Vahe and Dagdelen, John and Weston, Leigh and Dunn, Alexander and Rong, Ziqin and Kononova, Olga and Persson, Kristin A and Ceder, Gerbrand and Jain, Anubhav},
  journal={Nature},
  volume={571},
  number={7763},
  pages={95--98},
  year={2019},
  publisher={Nature Publishing Group},
  note={https://doi.org/10.1038/s41586-019-1335-8}  
}

@article{noh2015keyword,
  title={Keyword selection and processing strategy for applying text mining to patent analysis},
  author={Noh, Heeyong and Jo, Yeongran and Lee, Sungjoo},
  journal={Expert Syst. Appl.},
  volume={42},
  number={9},
  pages={4348--4360},
  year={2015},
  publisher={Elsevier},
  note={https://doi.org/10.1016/j.eswa.2015.01.050}  
}

@article{hu2017china,
  title={China as number one? {Evidence} from {China's} most recent patenting surge},
  author={Hu, Albert GZ and Zhang, Peng and Zhao, Lijing},
  journal={J. Dev. Econ.},
  volume={124},
  pages={107--119},
  year={2017},
  publisher={Elsevier},
  note={https://doi.org/10.1016/j.jdeveco.2016.09.004}  
}

@incollection{prud2019chinese,
  title     = {Chinese Patenting Trends and the Role of the State},
  author    = {Prud'homme, Dan and Zhang, Taolue},
  booktitle = {China’s Intellectual Property Regime for Innovation},
  pages     = {43--71},
  publisher = {Springer},
  address   = {Cham},
  year      = {2019}
}

@inproceedings{devlin2019bert,
  title     = {{BERT}: Pre-training of Deep Bidirectional Transformers for Language Understanding},
  author    = {Devlin, Jacob and Chang, Ming-Wei and Lee, Kenton and Toutanova, Kristina},
  editor    = {Burstein, Jill and Doran, Christy and Solorio, Thamar},
  booktitle = {Proceedings of the 2019 Conference of the North American Chapter of the Association for Computational Linguistics: Human Language Technologies, Volume 1 (Long and Short Papers)},
  address   = {Minneapolis, Minnesota},
  publisher = {Association for Computational Linguistics},
  pages     = {4171--4186},
  year      = {2019}
}

@article{reimers2019sentence,
  title={Sentence-{BERT}: Sentence embeddings using siamese {BERT}-networks},
  author={Reimers, Nils and Gurevych, Iryna},
  journal={arXiv preprint},
  year={2019},
  note={{,\space  https://arxiv.org/abs/1908.10084}}  
}

@inproceedings{song2020mpnet,
  title     = {{MPNet}: Masked and Permuted Pre-training for Language Understanding},
  author    = {Song, Kaitao and Tan, Xu and Qin, Tao and Lu, Jianfeng and Liu, Tie-Yan},
  editor    = {Larochelle, Hugo and Ranzato, Marc'Aurelio and Hadsell, Raia and Balcan, Maria-Florina and Lin, Hsuan-Tien},
  booktitle = {Advances in Neural Information Processing Systems},
  pages     = {16857--16867},
  publisher = {Curran Associates, Inc.},
  address   = {Red Hook, NY},
  year      = {2020}
}

@inproceedings{ni2022large,
  title     = {Large Dual Encoders Are Generalizable Retrievers},
  author    = {Ni, Jianmo and Qu, Chen and Lu, Jing and Dai, Zhuyun and Hernandez Abrego, Gustavo and Ma, Ji and Zhao, Vincent and Luan, Yi and Hall, Keith and Chang, Ming-Wei and Yang, Yinfei},
  editor    = {Goldberg, Yoav and Kozareva, Zornitsa and Zhang, Yue},
  booktitle = {Proceedings of the 2022 Conference on Empirical Methods in Natural Language Processing},
  address   = {Abu Dhabi, United Arab Emirates},
  publisher = {Association for Computational Linguistics},
  pages     = {9844--9855},
  year      = {2022}
}

@article{wang2022text,
  title={Text embeddings by weakly-supervised contrastive pre-training},
  author={Wang, Liang and Yang, Nan and Huang, Xiaolong and Jiao, Binxing and Yang, Linjun and Jiang, Daxin and Majumder, Rangan and Wei, Furu},
  journal={arXiv preprint},
  year={2022},
  note={{,\space  https://arxiv.org/abs/2212.03533}}  
}

@article{li2023towards,
  title={Towards general text embeddings with multi-stage contrastive learning},
  author={Li, Zehan and Zhang, Xin and Zhang, Yanzhao and Long, Dingkun and Xie, Pengjun and Zhang, Meishan},
  journal={arXiv preprint},
  year={2023},
  note={{,\space  https://arxiv.org/abs/2308.03281}}  
}

@inproceedings{le2014distributed,
  title     = {Distributed Representations of Sentences and Documents},
  author    = {Le, Quoc and Mikolov, Tomas},
  editor    = {Xing, Eric P. and Jebara, Tony},
  booktitle = {Proceedings of the 31st International Conference on Machine Learning},
  pages     = {1188--1196},
  address   = {Beijing, China},
  publisher = {PMLR},
  year      = {2014}
}

@inproceedings{vaswani2017attention,
  title     = {Attention Is All You Need},
  author    = {Vaswani, Ashish and Shazeer, Noam and Parmar, Niki and Uszkoreit, Jakob and Jones, Llion and Gomez, Aidan N. and Kaiser, {\L}ukasz and Polosukhin, Illia},
  editor    = {Guyon, Isabelle and von Luxburg, Ulrike and Bengio, Samy and Wallach, Hanna and Fergus, Rob and Vishwanathan, S. V. N. and Garnett, Roman},
  booktitle = {Advances in Neural Information Processing Systems},
  pages     = {5998--6008},
  publisher = {Curran Associates, Inc.},
  address   = {Red Hook, NY},
  year      = {2017}
}

@article{tian2023approximate,
  title={Approximate Nearest Neighbor Search in High Dimensional Vector Databases: Current Research and Future Directions},
  author={Tian, Yao and Yue, Ziyang and Zhang, Ruiyuan and Zhao, Xi and Zheng, Bolong and Zhou, Xiaofang},
  journal={IEEE Data Eng. Bull.},
  volume={46},
  number={3},
  pages={39--54},
  year={2023}
}

@misc{bernhardsson2017annoy,
  author       = {Bernhardsson, Erik},
  title        = {Annoy: Approximate Nearest Neighbors in {C++}/{Python} Optimized for Memory Usage and Loading/Saving to Disk},
  howpublished = {GitHub repository},
  year         = {2017},
  note         = {https://github.com/spotify/annoy (accessed 20 January 2026)}
}

@inproceedings{indyk1998approximate,
  title     = {Approximate Nearest Neighbors: Towards Removing the Curse of Dimensionality},
  author    = {Indyk, Piotr and Motwani, Rajeev},
  editor    = {Vitter, Jeffrey Scott},
  booktitle = {Proceedings of the Thirtieth Annual {ACM} Symposium on the Theory of Computing},
  address = {New York, NY, USA},
  publisher = {ACM},
  pages     = {604--613},
  year      = {1998}
}

@inproceedings{voorhees2000building,
  title     = {Building a Question Answering Test Collection},
  author    = {Voorhees, Ellen M. and Tice, Dawn M.},
  editor    = {Belkin, Nicholas J. and Ingwersen, Peter and Leong, Mun-Kew},
  booktitle = {Proceedings of the 23rd Annual International ACM SIGIR Conference on Research and Development in Information Retrieval},
  address   = {Athens, Greece},
  publisher = {ACM},
  pages     = {200--207},
  year      = {2000}
}

@article{jarvelin2002cumulated,
  title={Cumulated gain-based evaluation of {IR} techniques},
  author={J{\"a}rvelin, Kalervo and Kek{\"a}l{\"a}inen, Jaana},
  journal={ACM Trans. Inf. Syst.},
  volume={20},
  number={4},
  pages={422--446},
  year={2002},
  publisher={ACM New York, NY, USA},
  note={https://doi.org/10.1145/582415.582418}  
}

@misc{google_patents_result_viewer,
  author       = {{Google}},
  title        = {Result viewer},
  year         = {2026},
  note         = {https://support.google.com/faqs/answer/7049724 (accessed 20 January 2026)}
}






\end{document}